\documentclass[aps,pra,reprint,superscriptaddress,showpacs]{revtex4-1}

\usepackage{graphicx}
\usepackage{amsmath}
\usepackage{amsbsy}
\usepackage{float}
\usepackage{verbatim}
\usepackage{booktabs}
\usepackage{array}
\usepackage{tabularx}

\begin{document}
\newcolumntype{C}[1]{>{\centering\arraybackslash}m{#1}}


\title{Constructing a chiral dipolar mode in an achiral nanostructure}



\author{J\"org S.~Eismann}
\affiliation{Max Planck Institute for the Science of Light, Staudtstr. 2, D-91058 Erlangen, Germany}
\affiliation{Institute of Optics, Information and Photonics, University Erlangen-Nuremberg, Staudtstr. 7/B2, D-91058 Erlangen, Germany}
\author{Martin Neugebauer}
\affiliation{Max Planck Institute for the Science of Light, Staudtstr. 2, D-91058 Erlangen, Germany}
\affiliation{Institute of Optics, Information and Photonics, University Erlangen-Nuremberg, Staudtstr. 7/B2, D-91058 Erlangen, Germany}
\author{Peter Banzer}
\email[]{peter.banzer@mpl.mpg.de}
\homepage[]{http://www.mpl.mpg.de/}
\affiliation{Max Planck Institute for the Science of Light, Staudtstr. 2, D-91058 Erlangen, Germany}
\affiliation{Institute of Optics, Information and Photonics, University Erlangen-Nuremberg, Staudtstr. 7/B2, D-91058 Erlangen, Germany}


\date{\today}

\begin{abstract}
%
We discuss the excitation of a chiral dipolar mode in an achiral silicon nanoparticle. In particular, we make use of the electric and magnetic polarizabilities of the silicon nanoparticle to construct this chiral electromagnetic mode which is conceptually similar to the fundamental modes of 3D chiral nanostructures or molecules. We describe the chosen tailored excitation with a beam carrying neither spin nor orbital angular momentum and investigate the emission characteristics of the chiral dipolar mode in the helicity basis, consisting of parallel electric and magnetic dipole moments, phase shifted by $\pm \pi/2$. 
We demonstrate the wavelength dependence and measure the spin and orbital angular momentum in the emission of the excited chiral mode.
\end{abstract}


\maketitle

\section{Introduction}

The recent experimental demonstrations of electromagnetic dipolar modes induced in silicon and other high-refractive index nanoparticles highlight the importance of such structures as building blocks for novel nano-photonic devices and metasurfaces~\cite{Evlyukhin2012,Shi2012}. For example, the simultaneous and in-phase excitation of perpendicular electric and magnetic dipole moments --- commonly referred to as Huygens\textquoteright~dipole~\cite{Fu2013} --- leads to strongly directional scattering patterns, allowing for highly efficient routing and polarization multiplexing at the nanoscale~\cite{Staude2017}. Additionally, other combinations of electric and magnetic dipole moments can be achieved by providing carefully structured excitation fields~\cite{Wozniak2015, Neugebauer2016, Wei2017, Picardi2018, magspin}. 

In this letter, we investigate an electromagnetic dipole moment in the helicity basis~\cite{Zambrana-Puyalto2013}. Such a $\sigma$-dipole consists of parallel electric and magnetic dipole moments of equal amplitudes and a relative phase of $\pm \pi/2$, leading to a well-defined helicity of $\pm 1$ in the far field~\cite{Zambrana-Puyalto2013,Zambrana-Puyalto2016}. Furthermore, the $\sigma$-dipole is conceptually similar to the coupled dipole moments observed for the fundamental mode of a chiral nanostructure \cite{Barron2009, Gansel2009, Schäferling2014, Wozniak2018}. There, the occurence of a $\sigma$-dipole component is directly linked to the chiral geometry of the system. 
In our experiment, we excite the $\sigma$-dipole for the first time in an achiral nanoparticle with a tightly focused cylindrical vector beam, which is locally linearly polarized and carries neither spin nor orbital angular momentum. 
When tightly focused, the beam exhibits in-phase longitudinal electric and magnetic fields of equal amplitudes on the optical axis. However, because electric and magnetic polarizabilities of the particle are wavelength dependent, we obtain a superposition of phase delayed electric and magnetic longitudinal dipole moments induced in the particle, which results in the emission of light with predominately positive or negative helicity. 
Adapting the excitation wavelength, we maximize the helicity of the mode by balancing the dipole moments. The excited dipolar modes not only result in a far-field emission carrying spin angular momentum, but also orbital angular momentum occurs as a result of an azimuthal phase front. We begin with a theoretical comparison of a circularly polarized dipole and $\sigma$-dipoles. Then we describe our experimental scheme including the tailored excitation of a $\sigma$-dipole and finally, we compare experimental and theoretical results and elaborate on the far-field helicity and the chirality of the excited mode. 

\section{Circular and chiral dipolar modes}

%
%

\begin{figure*}
  \includegraphics[width=\textwidth]{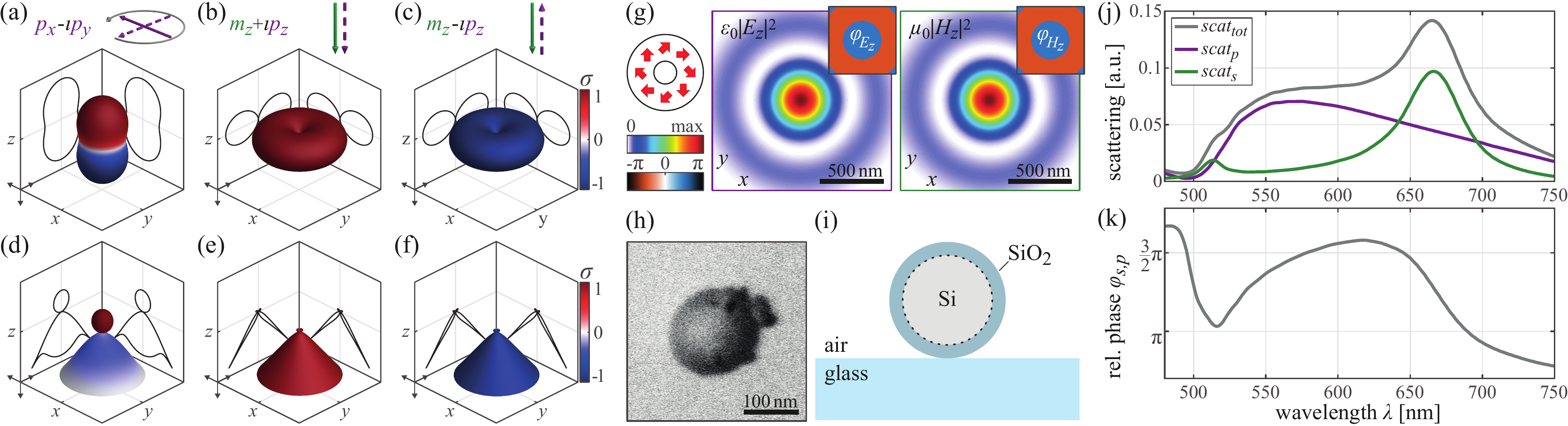}
  \caption{
(a)-(f) Radiation patterns of circularly polarized and $\sigma$-dipoles. The color-code corresponds to the helicity $\sigma$. The upper row shows the emission in free-space, the lower row shows the emission of the same dipoles placed 88\,nm above a glass substrate $(n_G = 1.53)$. 
(a), (d) Spinning dipole with dipole moments $\mathbf{p}= p_{0}\left(1,-\imath,0\right)$. (b), (e), (c), (f) $\sigma$-dipoles with dipole moments $\mathbf{p}= p_{0}\left(0,0,\pm \imath\right)$ and $\mathbf{m}= m_{0}\left(0,0,1\right)$.
(g) Incoming spiral polarization beam with locally linear polarization depicted by red arrows. The two field plots show the electric and magnetic $z$-component of the tightly focused beam in the focal plane. The phases are shown as insets.
(h) Scanning electron microscope image of the nanoparticle, used in the experiment. 
(i) Sketch of the system utilized for FDTD simulation. The particle has a crystalline silicon core with radius $r_{Si}=84$\,nm, a silicon-dioxide ($\text{SiO}_{2}$) shell of thickness $\delta=$4\,nm and is placed on an air-glass interface.
(j) Total scattering efficiency (gray line) retrieved from FDTD simulations. The purple and green lines depict the decomposition into $p$- and $s$-polarized components, corresponding to electric and magnetic multipoles respectively.
(k) Relative phase $\varphi_{s, p}$ between the scattered fields $E_s$ and $E_p$.
	}
  \label{fig:fig1}
\end{figure*}

In order to emphasize the fundamental difference between a spinning (circularly polarized) dipole and the $\sigma$-dipoles on which we will focus mainly in this study, we plot corresponding radiation patterns in Fig.~\ref{fig:fig1}~(a)-(f). The color-code refers to the helicity $\sigma$ defined by the normalized far-field Stokes parameter $S_3=(I_{rhc}-I_{lhc})/(I_{rhc}+I_{lhc})$~\cite{Jackson1999}, with $I_{rhc}$ and $I_{lhc}$ the intensities of the emitted right- and left-handed circularly polarized fields, respectively. In Fig.~\ref{fig:fig1}~(a)-(c) the dipoles are emitting in free-space, where (a) shows the circularly polarized electric dipole with $\mathbf{p}=p_0(1,-\imath,0)$. Fig.~\ref{fig:fig1} (b) and (c) depict two $\sigma$-dipoles with positive and negative helicities defined by a combination of $\mathbf{p}=p_0(0,0,\pm\imath)$ and $\mathbf{m}=m_0(0,0,1)$. 
Here $\mathbf{p}$ and $\mathbf{m}$ refer to the electric and magnetic dipole moments in Cartesian coordinates with $p_0=m_0/c_0$. For our geometry, where we only use $z$-oriented dipoles, the $\sigma$-dipoles can be expressed in a very compact form by using the dipole notation~\cite{Bliokh2014a}
\begin{equation}
\boldsymbol{\sigma}_\pm=A^{\pm} \left( \begin{matrix} \pm \imath \\ 1 \end{matrix}\right),
\label{eq:sigma}
\end{equation}
with $A^{\pm}$ the overall amplitudes of the dipoles.
For the regular spinning dipole in (a), the polarization of the emitted light is anisotropic with different signs of helicity in the upper and lower half-space~\cite{OConnor2014}. The average helicity is exactly zero. In contrast, for the two $\sigma$-dipoles in (b) and (c), the helicity is isotropic with an average of exactly $\pm1$~\cite{Zambrana-Puyalto2016}. In our experiment described below, the dipole emitter will be sitting on a dielectric interface for practical reasons. The dielectric interface near the emitter strongly alters the emission patterns (see~\cite{Lukosz1979} and references therein), enhancing the emission to the optically denser medium (here lower half-space with $z>0$).
The far-field emitted into the substrate can be calculated by using a plane-wave decomposition following ref.~\cite{Novotny2006}. 
Using the formalism introduced in Eq.~\eqref{eq:sigma}, for $\sigma$-dipoles it can be written as
\begin{align}\label{eqn:ff}
\mathbf{E}_{f}\left(k_{x},k_{y}\right)\propto
C\hat{\mathbf{T}}\left[ A^+ \left( \begin{matrix} -\imath \\ 1 \end{matrix}\right) + A^- \left( \begin{matrix} \imath \\ 1 \end{matrix}\right) \right]\text{,}
\end{align}
where $\mathbf{E}_{f}$ is expressed in the transverse magnetic ($E_{p}$) and transverse electric ($E_{s}$) polarization basis and $C=\frac{k_{\bot}}{k_0}\left[\left(k_{0}^{2}n_G^{2}-k_{\bot}^{2}\right)^{1/2}/k_{z}\right]\cdot\exp{\left(\imath k_{z}d\right)}$. Moreover, $k_0=2\pi / \lambda$ is the wave number in the upper half-space, $c_{0}$ the vacuum speed of light, $k_{\bot}=\left(k_{x}^2+k_{y}^2\right)^{1/2}$ the transverse wave number, $d$ is the distance of dipolar emitter to the interface and the matrix $\hat{\mathbf{T}}$ contains the Fresnel transmission coefficients $t_{s}$ and $t_{p}$~\cite{Novotny2006}
\begin{align}
\hat{\mathbf{T}}=& \left(\begin{matrix} t_{p}&0\\ 0&t_{s} \end{matrix}\right)\text{.}
\end{align}
In Fig.~\ref{fig:fig1}~(d)-(f) we show the far fields of the aforementioned dipoles but now positioned $d=88$\,nm above an air-glass interface with refractive index of the glass substrate $n_G=1.53$, adapted to the experiment described later. Because of the substrate, the average helicity does not reach values of $\pm1$ (in contrast to free-space), but is typically slightly lower because of the Fresnel coefficients breaking the dual symmetry of the system~\cite{Zambrana-Puyalto2016}. 

\section{Tailored excitation scheme}

Our intention is the tailored excitation and verification of $\sigma$-dipole moments inside a spherical silicon particle. For this purpose, we first design an excitation field capable of inducing electric and magnetic dipole moments, which are oriented parallel and have a phase difference of $\pi/2$. For paraxial beams this is not possible because there electric and magnetic fields are always perpendicular to each other \cite{Jackson1999}. One possible solution can be realized by tightly focusing a superposition of radially and azimuthally polarized vector beams, which exhibit purely $z$-polarized electric and magnetic fields on the optical axis, respectively \cite{Wozniak2015, Youngworth2000}. To achieve the desired relative phase between the electric and magnetic dipole moments, the inherent properties of the silicon particle can be utilized. Due to the complex electric and magnetic polarizabilities of the particle, a wavelength dependent phase shift between the fields and the resulting dipolar modes is introduced~\cite{Neugebauer2016}. 
Accordingly, for our experimental implementation, we utilize a superposition of radially and azimuthally polarized beams of equal amplitudes --- a so-called spiral polarization beam --- and focus it tightly. In Fig.~\ref{fig:fig1} (g), left, the local polarization vector of the incoming paraxial beam before focusing is sketched with red arrows. The calculated focal field distributions of the relevant $z$-components are depicted on the right, with the corresponding phases plotted as insets. On the optical axis, the electric and the magnetic fields are parallel, $z$-polarized and in-phase. All other field components are zero on-axis. Accordingly the induced dipole moments at this position can be written as $p_z=\alpha^e E_z$ and $m_z=\alpha^m H_z$,
where $\alpha^{e}$ and $\alpha^{m}$ are complex polarizabilities, describing the excitation of the corresponding dipoles by electromagnetic fields~\cite{Bliokh2014a}. It is important to emphasize the absence of any magnetoelectric or chiral terms in the polarizability, which describe a coupling of electric and magnetic modes. The symmetries of a spherical structure do not allow for such interaction terms, which would be present for a chiral nanostructure or a bianisotropic particle~\cite{Bliokh2014a,Alaee2015}. 

Next, we discuss the optical response of the chosen silicon nanoparticle, similar to those used in~\cite{Neugebauer2016, Wozniak2015}. A scanning electron microscope image of the particle is shown in Fig.~\ref{fig:fig1}~(h). We run finite difference time-domain (FDTD) simulations using the same beam parameters and far-field collection geometry as in the experiment to investigate the spectral response of the silicon nano-sphere with core radius $r_{Si}=84$\,nm, surrounded by a silicon-dioxide shell of estimated thickness $\delta=4$\,nm, sitting on a glass substrate [see Fig.~\ref{fig:fig1}~(i)]. The wavelength dependent scattering efficiencies, decomposed into contributions of $E_p$ and $E_s$ are depicted in Fig.~\ref{fig:fig1}~(j) as purple and green lines respectively. Looking at Eq.~\ref{eqn:ff}, we see that for the chosen geometry, the far field of an electric $p_z$ dipole is purely $p$-polarized, while for the magnetic $m_z$ dipole it is purely $s$-polarized. Also the magnetic quadrupole component which can be excited in our system features a purely $s$-polarized far field~\cite{Wozniak2015}. Hence, we can associate the purple curve with the electric dipole and the stronger resonance of the green curve with the magnetic dipole. The second substantially weaker resonance of the green curve corresponds to the magnetic quadrupole. Since we want to excite a $\sigma$-dipole, the phase difference between the $s$- and $p$-polarized field components is of special interest for us. For that reason we plot the relative phase between $E_s$ and $E_p$ in Fig.~\ref{fig:fig1}~(k), evaluated from the simulated data in the angular region below the critical angle. We see that in the range of 630 to 710\,nm the relative phase depends strongly on the wavelength of the incoming beam. This provides a large tuning range of the excited dipolar mode. Also in this region the contribution of the magnetic quadrupole with its response peaking at 510\,nm is completely negligible.

\section{Implementation }
\label{sec:Implementation}

\begin{figure} 
  \includegraphics[width=0.48\textwidth]{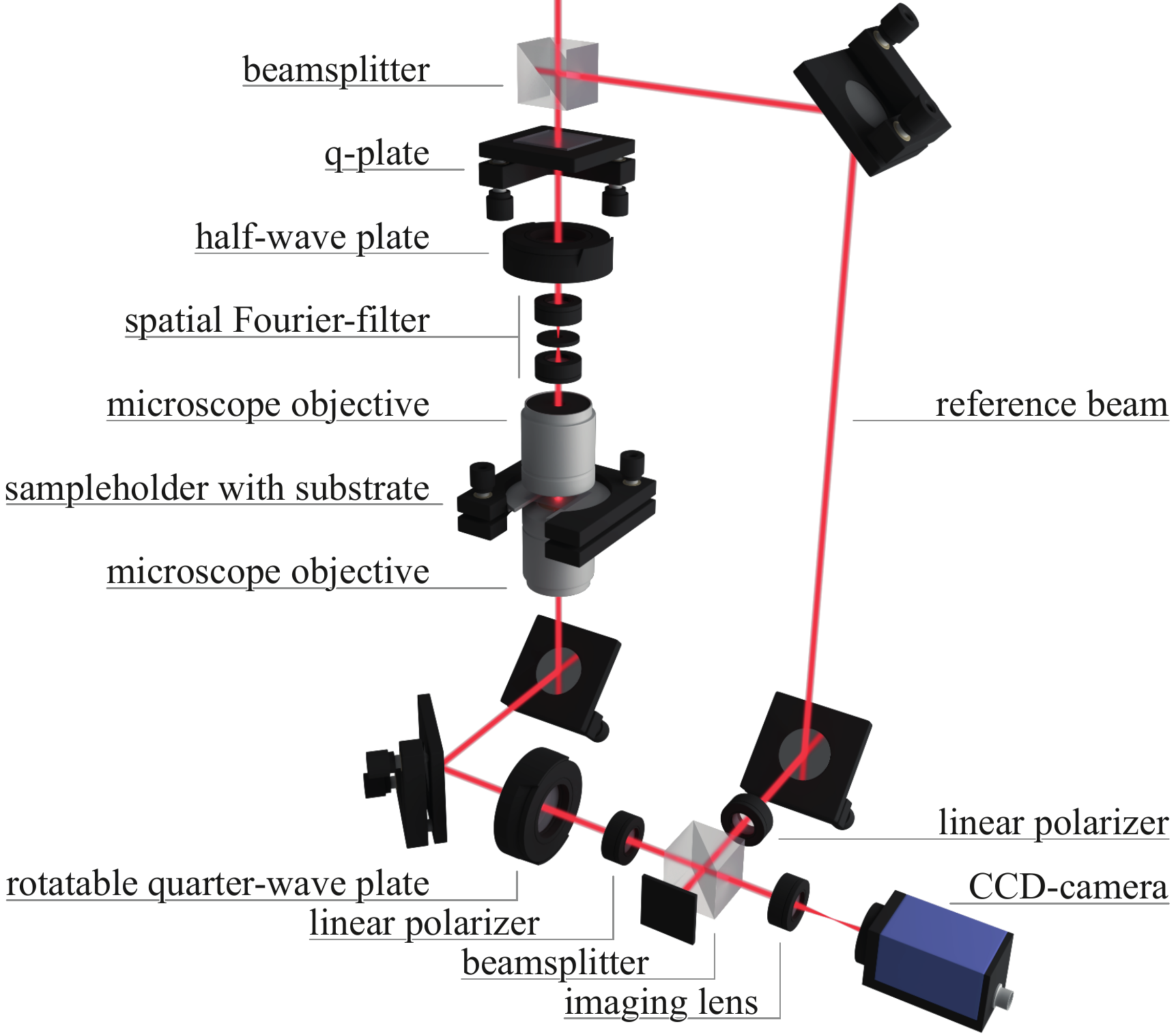} 
  \caption{Schematic draft of the measurement setup. The incoming linearly polarized Gaussian beam is split into two parts. The main beam is tailored into the desired spiral polarization beam and afterwards focused on a silicon nanostructure by a microscope objective with a numerical aperture (NA) of 0.9. A second microscope objective collects the incoming beam as well as the scattered light up to an NA of 1.3. Using a rotatable quarter-wave plate, a full back focal plane Stokes analysis is implemented. Additionally, the reference beam can be superimposed for interferometric measurements. 
}
  \label{fig:setup}
\end{figure}
As a light source for our measurement, a super continuum laser source was utilized with an acousto-optical tunable filter to select the desired wavelength between 480\,nm and 710\,nm with a bandwidth of approximately 2\,nm. The output of the filter is guided by a singlemode fiber, which provides a collimated, linearly-polarized Gaussian beam at the entrance of the main setup, shown in a simplified version in Fig.~\ref{fig:setup}~\cite{Banzer2010}. 
In order to verify the generation of OAM by interferometric measurements, a beam splitter is used to split the incoming beam into two parts.
The main beam is converted by a q-plate~\cite{Marrucci2006} of charge $-\frac{1}{2}$ and a half-wave plate into the desired spiral polarization beam, which is spatially Fourier-filtered afterwards. A microscope objective (MO) with high numerical aperture (NA) of 0.9 is used for tightly focusing the beam onto a high-refractive-index silicon nano-sphere, placed on a glass substrate \cite{Mick2014}.
The glass substrate is mounted on a sample holder, attached to a 3D-piezo-stage allowing for precise positioning of the particle with respect to the beam. The collection of the transmitted and forward scattered light is realized by an oil-immersion-type MO with an NA of 1.3. The two microscope objectives form a confocal alignment, to provide a collimated output of the second MO. 
We utilize a rotatable broadband quarter-wave plate in combination with a fixed linear polarizer, which projects the light onto different polarization states in order to perform a full Stokes parameter measurement \cite{Schaefer2007}. Thereafter, the beam passes a second beam-splitter, which can be used for phase measurements by overlapping the transmitted light with the reference beam. Finally, we image the back focal plane (BFP) of the lower MO onto the CCD camera. Because the collecting MO has a higher NA than the MO used for focusing of the beam, in the angular range defined by $\text{NA}=\left[0.9, 1.3\right]$ no contribution of the excitation field itself is found, but only scattered light. This allows for investigating the type of the excited electromagnetic dipole. For the interference measurements, the linearly polarized reference beam was overlaped with the scattered light under a small angle.

\section{Experimental results and discussion}

\begin{figure*} 
  \includegraphics[width=\textwidth]{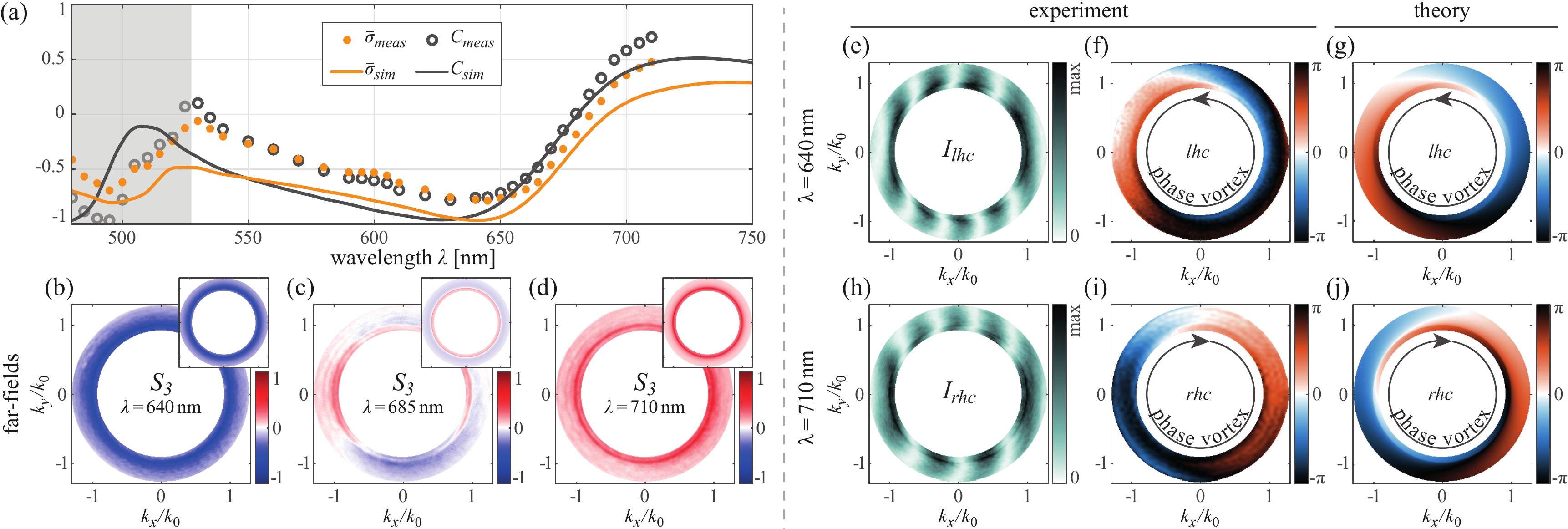}
  \caption{(a) Spectral measurement of average helicity $\bar{\sigma}$ and chirality $C$ depicted as orange dots and gray circles, respectively. The solid lines show corresponding results achieved by FDTD simulation. The region where the magnetic quadrupole should be taken into account for a more accurate evaluation of the chirality is shaded gray.
(b)-(d) Measured BFP images showing $S_3$ for three different wavelengths, where $\bar{\sigma}$ is maximum, minimum and close to zero. The images computed by fitting the far fields to the measured data are shown as insets.
(e) Interference pattern of left-handed circular polarized field, measured at 640\,nm with (f) reconstructed experimental phase of the left-handed circularly (lhc) polarized field component and (g) associated theoretical phase distribution for lhc field component.
(h)-(j) Corresponding images for right-handed circular (rhc) polarization at 710\,nm.
	}
  \label{fig:meas1}
\end{figure*}

For the first measurement we investigated the spectral behavior of the average helicity $ \bar{\sigma}= (W_{rhc}-W_{lhc})/(W_{rhc}+W_{lhc})$ \cite{Bekshaev2011a}, 
with $W_{rhc}$ and $W_{lhc}$ the integrated right- and left-handed circularly polarized intensities in the angular range defined by $\text{NA}=\left[0.92, 1.28\right]$. This corresponds to an average of the normalized Stokes parameter $S_3$. The measurement results are shown as orange dots in Fig.~\ref{fig:meas1} (a). In addition, the corresponding curve retrieved from the aforementioned FDTD simulation is shown as an orange line. 
Numerical and experimental data are in good agreement. In Fig.~\ref{fig:meas1} (b)-(d), we display the experimentally measured far-field $S_{3}$ parameter for three selected wavelengths, where the measured average helicity is maximum, minimum and close to zero. 
In order to determine the excited dipolar mode, we fit the far-field Stokes parameters of the $\sigma$-dipoles, calculated with Eq.~\eqref{eqn:ff}, to our measured data and use their complex amplitudes as free parameters. 
The theoretical far-fields obtained by this method are displayed as insets in Fig.~\ref{fig:meas1} (b)-(d). As an estimation for the quality of the fit, we compute the overlap integral of the fitted and the experimental electric fields which turns out to be above $96\%$ for all three wavelengths. Adapting the formalism from \cite{Zambrana-Puyalto2016} we calculate the chirality of the measured dipole moments:
\begin{equation}
C(A^{\pm})=\frac{|A^+|^2-|A^-|^2}{|A^+|^2+|A^-|^2}.
\label{eq:chirality}
\end{equation}
The absolute squared dipole amplitudes, as well as the calculated field overlap and the chirality of the three dipolar modes are listed in Tab.~\ref{tab:tab}. 
Using this approach, we also determined the chirality for all measured wavelengths and show it in Fig.~\ref{fig:meas1} (a) as gray circles, with a corresponding gray curve obtained by FDTD simulation. The shaded area indicates the spectral range below 530\,nm, where the contribution of the magnetic quadrupole is not negligible anymore. This should be taken into account for a more precise evaluation of the chirality, where higher order multipoles are included in the analysis. Therefore, we consider the chirality to be accurate only above 530\,nm. 

Our experimentally reconstructed dipole moments are not exclusively $\sigma_+$ or $\sigma_-$. However our measurements show that we are able to excite a dipolar mode, which is strongly dominated by one or the other $\sigma$-dipole moment. In addition we see that within a spectral range of only 70\,nm it is possible to tune the excited dipole between two modes of opposite dominating helicity (+ or -). The two modes of maximum chirality exhibit 88\,\% of $\sigma_-$ and 85\,\% of $\sigma_+$ for 640\,nm and 710\,nm, respectively.

\begin{table} 
	\caption{Results of the calculated parameters.}\label{tab:tab}
	\begin{ruledtabular}
	\begin{tabular}{>{\centering}m{0.18\columnwidth} C{0.18\columnwidth} C{0.18\columnwidth} C{0.18\columnwidth} C{0.18\columnwidth}}
	Wavelength $[$nm$]$ & Field-overlap & $|A^+|^2$ & $|A^-|^2$ & chirality ~$C$~\\
	\hline 
	\vspace{0.1cm} 640 & \vspace{0.1cm} 0.98 & \vspace{0.1cm} 0.12 & \vspace{0.1cm} 0.88 & -0.76 \\
	685 & 0.96 & 0.59 & 0.41 & ~0.18\\
	710 & 0.97 & 0.85 & 0.15 & ~0.70\\
	\end{tabular} \end{ruledtabular} 
\end{table}

As a last step to verify the excitation of $\sigma$-dipoles, we examine the conservation of angular momentum in our system. As already mentioned, our incoming beam is purely linearly polarized and hence, carries zero spin angular momentum (SAM). Furthermore, it features a planar phase front and therefore has also zero orbital angular momentum (OAM). Still, Fig.~\ref{fig:meas1}~(a)-(d) clearly indicate the generation of circularly polarized fields scattered in forward direction. For the purpose of angular momentum conservation in our cylindrically symmetric system~\cite{Dogariu2006,Bliokh2014b,Bliokh2015}, the emitted light therefore must feature also OAM of the opposite sign, which should appear as a phase-vortex of $\pm2\pi$ in the scattered light. These phase vortices of charge $\pm1$, together with the polarization resolved far-fields allow for an unambiguous identification of the $\sigma$-dipoles. The measured interference patterns for the left- and right-handed circular polarization components are shown in Fig.~\ref{fig:meas1} (e) and (h). 
Counting the fringes in the upper and lower half of the images, we see that for lhc-polarization there is one additional fringe at the lower half and for rhc-polarization there is one additional fringe at the upper half, indicating opposite signs of OAM. A full phase reconstruction following the technique explained in \cite{Takeda1982} results in the measured phase distributions shown in Fig.~\ref{fig:meas1}~(f) and (i). The experimental phase images are in very good agreement with the theoretically calculated phase distributions [see Fig.~\ref{fig:meas1}~(g) and (j)]. In particular, they prove the generation of the two phase vortices of oposite signs indicating the excitation of $\sigma$-dipole moments. 

%
%

\section{Summary}

We used structured illumination to excite $\sigma$-dipole moments in a spherical achiral dielectric nanoparticle. Furthermore, we verified these modes by measuring their far-field helicity and orbital angular momentum and investigated their spectral dependence. Hence, the interaction of the chosen tightly focused spiral polarization beam with an achiral nanoparticle (silicon nanosphere) placed on-axis in the focal plane allows for the excitation of $\sigma$-dipoles, resulting in the emission of structured light carrying both SAM and OAM. Our study highlights the importance of structured light in nanoscale light-matter interactions. Furthermore, it sheds new light on the definition of chiral dipoles known from 3D chiral nanostructures or molecules. For such chiral building-blocks, the geometry causes the excitation of a chiral dipole, whereas in our case it is the excitation field ruling the interaction.

\begin{acknowledgments}
We gratefully acknowledge fruitful discussions with Sergey Nechayev and Gerd Leuchs.
\end{acknowledgments}

%
\bibliography{bib}

\begin{thebibliography}{31}%
\makeatletter
\providecommand \@ifxundefined [1]{%
 \@ifx{#1\undefined}
}%
\providecommand \@ifnum [1]{%
 \ifnum #1\expandafter \@firstoftwo
 \else \expandafter \@secondoftwo
 \fi
}%
\providecommand \@ifx [1]{%
 \ifx #1\expandafter \@firstoftwo
 \else \expandafter \@secondoftwo
 \fi
}%
\providecommand \natexlab [1]{#1}%
\providecommand \enquote  [1]{``#1''}%
\providecommand \bibnamefont  [1]{#1}%
\providecommand \bibfnamefont [1]{#1}%
\providecommand \citenamefont [1]{#1}%
\providecommand \href@noop [0]{\@secondoftwo}%
\providecommand \href [0]{\begingroup \@sanitize@url \@href}%
\providecommand \@href[1]{\@@startlink{#1}\@@href}%
\providecommand \@@href[1]{\endgroup#1\@@endlink}%
\providecommand \@sanitize@url [0]{\catcode `\\12\catcode `\$12\catcode
  `\&12\catcode `\#12\catcode `\^12\catcode `\_12\catcode `\%12\relax}%
\providecommand \@@startlink[1]{}%
\providecommand \@@endlink[0]{}%
\providecommand \url  [0]{\begingroup\@sanitize@url \@url }%
\providecommand \@url [1]{\endgroup\@href {#1}{\urlprefix }}%
\providecommand \urlprefix  [0]{URL }%
\providecommand \Eprint [0]{\href }%
\providecommand \doibase [0]{http://dx.doi.org/}%
\providecommand \selectlanguage [0]{\@gobble}%
\providecommand \bibinfo  [0]{\@secondoftwo}%
\providecommand \bibfield  [0]{\@secondoftwo}%
\providecommand \translation [1]{[#1]}%
\providecommand \BibitemOpen [0]{}%
\providecommand \bibitemStop [0]{}%
\providecommand \bibitemNoStop [0]{.\EOS\space}%
\providecommand \EOS [0]{\spacefactor3000\relax}%
\providecommand \BibitemShut  [1]{\csname bibitem#1\endcsname}%
\let\auto@bib@innerbib\@empty
\bibitem [{\citenamefont {Evlyukhin}\ \emph {et~al.}(2012)\citenamefont
  {Evlyukhin}, \citenamefont {Novikov}, \citenamefont {Zywietz}, \citenamefont
  {Eriksen}, \citenamefont {Reinhardt}, \citenamefont {Bozhevolnyi},\ and\
  \citenamefont {Chichkov}}]{Evlyukhin2012}%
  \BibitemOpen
  \bibfield  {author} {\bibinfo {author} {\bibfnamefont {A.~B.}\ \bibnamefont
  {Evlyukhin}}, \bibinfo {author} {\bibfnamefont {S.~M.}\ \bibnamefont
  {Novikov}}, \bibinfo {author} {\bibfnamefont {U.}~\bibnamefont {Zywietz}},
  \bibinfo {author} {\bibfnamefont {R.~L.}\ \bibnamefont {Eriksen}}, \bibinfo
  {author} {\bibfnamefont {C.}~\bibnamefont {Reinhardt}}, \bibinfo {author}
  {\bibfnamefont {S.~I.}\ \bibnamefont {Bozhevolnyi}}, \ and\ \bibinfo {author}
  {\bibfnamefont {B.~N.}\ \bibnamefont {Chichkov}},\ }\href {\doibase
  10.1021/nl301594s} {\bibfield  {journal} {\bibinfo  {journal} {Nano Lett.}\
  }\textbf {\bibinfo {volume} {12}},\ \bibinfo {pages} {3749} (\bibinfo {year}
  {2012})}\BibitemShut {NoStop}%
\bibitem [{\citenamefont {Shi}\ \emph {et~al.}(2012)\citenamefont {Shi},
  \citenamefont {Tuzer}, \citenamefont {Fenollosa},\ and\ \citenamefont
  {Meseguer}}]{Shi2012}%
  \BibitemOpen
  \bibfield  {author} {\bibinfo {author} {\bibfnamefont {L.}~\bibnamefont
  {Shi}}, \bibinfo {author} {\bibfnamefont {T.~U.}\ \bibnamefont {Tuzer}},
  \bibinfo {author} {\bibfnamefont {R.}~\bibnamefont {Fenollosa}}, \ and\
  \bibinfo {author} {\bibfnamefont {F.}~\bibnamefont {Meseguer}},\ }\href
  {\doibase 10.1002/adma.201201987} {\bibfield  {journal} {\bibinfo  {journal}
  {Adv. Mater.}\ }\textbf {\bibinfo {volume} {24}},\ \bibinfo {pages} {5934}
  (\bibinfo {year} {2012})}\BibitemShut {NoStop}%
\bibitem [{\citenamefont {Fu}\ \emph {et~al.}(2013)\citenamefont {Fu},
  \citenamefont {Kuznetsov}, \citenamefont {Miroshnichenko}, \citenamefont
  {Yu},\ and\ \citenamefont {Luk'yanchuk}}]{Fu2013}%
  \BibitemOpen
  \bibfield  {author} {\bibinfo {author} {\bibfnamefont {Y.~H.}\ \bibnamefont
  {Fu}}, \bibinfo {author} {\bibfnamefont {A.~I.}\ \bibnamefont {Kuznetsov}},
  \bibinfo {author} {\bibfnamefont {A.~E.}\ \bibnamefont {Miroshnichenko}},
  \bibinfo {author} {\bibfnamefont {Y.~F.}\ \bibnamefont {Yu}}, \ and\ \bibinfo
  {author} {\bibfnamefont {B.}~\bibnamefont {Luk'yanchuk}},\ }\href {\doibase
  10.1038/ncomms2538} {\bibfield  {journal} {\bibinfo  {journal} {Nat.
  Commun.}\ }\textbf {\bibinfo {volume} {4}},\ \bibinfo {pages} {1527}
  (\bibinfo {year} {2013})}\BibitemShut {NoStop}%
\bibitem [{\citenamefont {Staude}\ and\ \citenamefont
  {Schilling}(2017)}]{Staude2017}%
  \BibitemOpen
  \bibfield  {author} {\bibinfo {author} {\bibfnamefont {I.}~\bibnamefont
  {Staude}}\ and\ \bibinfo {author} {\bibfnamefont {J.}~\bibnamefont
  {Schilling}},\ }\href {\doibase 10.1038/nphoton.2017.39} {\bibfield
  {journal} {\bibinfo  {journal} {Nat. Photon.}\ }\textbf {\bibinfo {volume}
  {11}},\ \bibinfo {pages} {274} (\bibinfo {year} {2017})}\BibitemShut
  {NoStop}%
\bibitem [{\citenamefont {Wo{\'{z}}niak}\ \emph {et~al.}(2015)\citenamefont
  {Wo{\'{z}}niak}, \citenamefont {Banzer},\ and\ \citenamefont
  {Leuchs}}]{Wozniak2015}%
  \BibitemOpen
  \bibfield  {author} {\bibinfo {author} {\bibfnamefont {P.}~\bibnamefont
  {Wo{\'{z}}niak}}, \bibinfo {author} {\bibfnamefont {P.}~\bibnamefont
  {Banzer}}, \ and\ \bibinfo {author} {\bibfnamefont {G.}~\bibnamefont
  {Leuchs}},\ }\href {\doibase 10.1002/lpor.201400188} {\bibfield  {journal}
  {\bibinfo  {journal} {Laser Photon. Rev.}\ }\textbf {\bibinfo {volume} {9}},\
  \bibinfo {pages} {231} (\bibinfo {year} {2015})}\BibitemShut {NoStop}%
\bibitem [{\citenamefont {Neugebauer}\ \emph {et~al.}(2016)\citenamefont
  {Neugebauer}, \citenamefont {Wo{\'{z}}niak}, \citenamefont {Bag},
  \citenamefont {Leuchs},\ and\ \citenamefont {Banzer}}]{Neugebauer2016}%
  \BibitemOpen
  \bibfield  {author} {\bibinfo {author} {\bibfnamefont {M.}~\bibnamefont
  {Neugebauer}}, \bibinfo {author} {\bibfnamefont {P.}~\bibnamefont
  {Wo{\'{z}}niak}}, \bibinfo {author} {\bibfnamefont {A.}~\bibnamefont {Bag}},
  \bibinfo {author} {\bibfnamefont {G.}~\bibnamefont {Leuchs}}, \ and\ \bibinfo
  {author} {\bibfnamefont {P.}~\bibnamefont {Banzer}},\ }\href {\doibase
  10.1038/ncomms11286 OPEN} {\bibfield  {journal} {\bibinfo  {journal} {Nat.
  Commun.}\ }\textbf {\bibinfo {volume} {7}},\ \bibinfo {pages} {11286}
  (\bibinfo {year} {2016})}\BibitemShut {NoStop}%
\bibitem [{\citenamefont {Wei}\ \emph {et~al.}(2017)\citenamefont {Wei},
  \citenamefont {Bhattacharya},\ and\ \citenamefont {{Paul Urbach}}}]{Wei2017}%
  \BibitemOpen
  \bibfield  {author} {\bibinfo {author} {\bibfnamefont {L.}~\bibnamefont
  {Wei}}, \bibinfo {author} {\bibfnamefont {N.}~\bibnamefont {Bhattacharya}}, \
  and\ \bibinfo {author} {\bibfnamefont {H.}~\bibnamefont {{Paul Urbach}}},\
  }\href {\doibase 10.1364/OL.42.001776} {\bibfield  {journal} {\bibinfo
  {journal} {Opt. Lett.}\ }\textbf {\bibinfo {volume} {42}},\ \bibinfo {pages}
  {1776} (\bibinfo {year} {2017})},\ \Eprint {http://arxiv.org/abs/1611.04351}
  {arXiv:1611.04351} \BibitemShut {NoStop}%
\bibitem [{\citenamefont {Picardi}\ \emph {et~al.}(2018)\citenamefont
  {Picardi}, \citenamefont {Zayats},\ and\ \citenamefont
  {Rodr{\'{i}}guez-Fortu{\~{n}}o}}]{Picardi2018}%
  \BibitemOpen
  \bibfield  {author} {\bibinfo {author} {\bibfnamefont {M.~F.}\ \bibnamefont
  {Picardi}}, \bibinfo {author} {\bibfnamefont {A.~V.}\ \bibnamefont {Zayats}},
  \ and\ \bibinfo {author} {\bibfnamefont {F.~J.}\ \bibnamefont
  {Rodr{\'{i}}guez-Fortu{\~{n}}o}},\ }\href {\doibase
  10.1103/PhysRevLett.120.117402} {\bibfield  {journal} {\bibinfo  {journal}
  {Phys. Rev. Lett.}\ }\textbf {\bibinfo {volume} {120}},\ \bibinfo {pages}
  {117402} (\bibinfo {year} {2018})}\BibitemShut {NoStop}%
\bibitem [{\citenamefont {{Neugebauer}}\ \emph {et~al.}(2017)\citenamefont
  {{Neugebauer}}, \citenamefont {{Eismann}}, \citenamefont {{Bauer}},\ and\
  \citenamefont {{Banzer}}}]{magspin}%
  \BibitemOpen
  \bibfield  {author} {\bibinfo {author} {\bibfnamefont {M.}~\bibnamefont
  {{Neugebauer}}}, \bibinfo {author} {\bibfnamefont {J.}~\bibnamefont
  {{Eismann}}}, \bibinfo {author} {\bibfnamefont {T.}~\bibnamefont {{Bauer}}},
  \ and\ \bibinfo {author} {\bibfnamefont {P.}~\bibnamefont {{Banzer}}},\
  }\href@noop {} {\bibfield  {journal} {\bibinfo  {journal} {ArXiv e-prints}\ }
  (\bibinfo {year} {2017})},\ \Eprint {http://arxiv.org/abs/1711.10268}
  {arXiv:1711.10268 [physics.optics]} \BibitemShut {NoStop}%
\bibitem [{\citenamefont {Zambrana-Puyalto}\ \emph {et~al.}(2013)\citenamefont
  {Zambrana-Puyalto}, \citenamefont {Fernandez-Corbaton}, \citenamefont {Juan},
  \citenamefont {Vidal},\ and\ \citenamefont
  {Molina-Terriza}}]{Zambrana-Puyalto2013}%
  \BibitemOpen
  \bibfield  {author} {\bibinfo {author} {\bibfnamefont {X.}~\bibnamefont
  {Zambrana-Puyalto}}, \bibinfo {author} {\bibfnamefont {I.}~\bibnamefont
  {Fernandez-Corbaton}}, \bibinfo {author} {\bibfnamefont {M.~L.}\ \bibnamefont
  {Juan}}, \bibinfo {author} {\bibfnamefont {X.}~\bibnamefont {Vidal}}, \ and\
  \bibinfo {author} {\bibfnamefont {G.}~\bibnamefont {Molina-Terriza}},\ }\href
  {\doibase 10.1364/OL.38.001857} {\bibfield  {journal} {\bibinfo  {journal}
  {Opt. Lett.}\ }\textbf {\bibinfo {volume} {38}},\ \bibinfo {pages} {1857}
  (\bibinfo {year} {2013})}\BibitemShut {NoStop}%
\bibitem [{\citenamefont {Zambrana-Puyalto}\ and\ \citenamefont
  {Bonod}(2016)}]{Zambrana-Puyalto2016}%
  \BibitemOpen
  \bibfield  {author} {\bibinfo {author} {\bibfnamefont {X.}~\bibnamefont
  {Zambrana-Puyalto}}\ and\ \bibinfo {author} {\bibfnamefont {N.}~\bibnamefont
  {Bonod}},\ }\href {\doibase 10.1039/C6NR00676K} {\bibfield  {journal}
  {\bibinfo  {journal} {Nanoscale}\ }\textbf {\bibinfo {volume} {8}},\ \bibinfo
  {pages} {10441} (\bibinfo {year} {2016})}\BibitemShut {NoStop}%
\bibitem [{\citenamefont {Barron}(2009)}]{Barron2009}%
  \BibitemOpen
  \bibfield  {author} {\bibinfo {author} {\bibfnamefont {L.~D.}\ \bibnamefont
  {Barron}},\ }\enquote {\bibinfo {title} {An introduction to chirality at the
  nanoscale},}\ in\ \href {\doibase 10.1002/9783527625345.ch1} {\emph {\bibinfo
  {booktitle} {Chirality at the Nanoscale}}}\ (\bibinfo  {publisher}
  {Wiley-Blackwell},\ \bibinfo {year} {2009})\ Chap.~\bibinfo {chapter} {1},
  pp.\ \bibinfo {pages} {1--27}\BibitemShut {NoStop}%
\bibitem [{\citenamefont {Gansel}\ \emph {et~al.}(2009)\citenamefont {Gansel},
  \citenamefont {Thiel}, \citenamefont {Rill}, \citenamefont {Decker},
  \citenamefont {Bade}, \citenamefont {Saile}, \citenamefont {von Freymann},
  \citenamefont {Linden},\ and\ \citenamefont {Wegener}}]{Gansel2009}%
  \BibitemOpen
  \bibfield  {author} {\bibinfo {author} {\bibfnamefont {J.~K.}\ \bibnamefont
  {Gansel}}, \bibinfo {author} {\bibfnamefont {M.}~\bibnamefont {Thiel}},
  \bibinfo {author} {\bibfnamefont {M.~S.}\ \bibnamefont {Rill}}, \bibinfo
  {author} {\bibfnamefont {M.}~\bibnamefont {Decker}}, \bibinfo {author}
  {\bibfnamefont {K.}~\bibnamefont {Bade}}, \bibinfo {author} {\bibfnamefont
  {V.}~\bibnamefont {Saile}}, \bibinfo {author} {\bibfnamefont
  {G.}~\bibnamefont {von Freymann}}, \bibinfo {author} {\bibfnamefont
  {S.}~\bibnamefont {Linden}}, \ and\ \bibinfo {author} {\bibfnamefont
  {M.}~\bibnamefont {Wegener}},\ }\href {\doibase 10.1126/science.1177031}
  {\bibfield  {journal} {\bibinfo  {journal} {Science}\ }\textbf {\bibinfo
  {volume} {325}},\ \bibinfo {pages} {1513} (\bibinfo {year}
  {2009})}\BibitemShut {NoStop}%
\bibitem [{\citenamefont {Schäferling}\ \emph {et~al.}(2014)\citenamefont
  {Schäferling}, \citenamefont {Yin}, \citenamefont {Engheta},\ and\
  \citenamefont {Giessen}}]{Schäferling2014}%
  \BibitemOpen
  \bibfield  {author} {\bibinfo {author} {\bibfnamefont {M.}~\bibnamefont
  {Schäferling}}, \bibinfo {author} {\bibfnamefont {X.}~\bibnamefont {Yin}},
  \bibinfo {author} {\bibfnamefont {N.}~\bibnamefont {Engheta}}, \ and\
  \bibinfo {author} {\bibfnamefont {H.}~\bibnamefont {Giessen}},\ }\href
  {\doibase 10.1021/ph5000743} {\bibfield  {journal} {\bibinfo  {journal} {ACS
  Photonics}\ }\textbf {\bibinfo {volume} {1}},\ \bibinfo {pages} {530}
  (\bibinfo {year} {2014})}\BibitemShut {NoStop}%
\bibitem [{\citenamefont {Wo{\'{z}}niak}\ \emph {et~al.}(2018)\citenamefont
  {Wo{\'{z}}niak}, \citenamefont {De~Leon}, \citenamefont {Höflich},
  \citenamefont {Haverkamp}, \citenamefont {Christiansen}, \citenamefont
  {Leuchs},\ and\ \citenamefont {Banzer}}]{Wozniak2018}%
  \BibitemOpen
  \bibfield  {author} {\bibinfo {author} {\bibfnamefont {P.}~\bibnamefont
  {Wo{\'{z}}niak}}, \bibinfo {author} {\bibfnamefont {I.}~\bibnamefont
  {De~Leon}}, \bibinfo {author} {\bibfnamefont {K.}~\bibnamefont {Höflich}},
  \bibinfo {author} {\bibfnamefont {C.}~\bibnamefont {Haverkamp}}, \bibinfo
  {author} {\bibfnamefont {S.}~\bibnamefont {Christiansen}}, \bibinfo {author}
  {\bibfnamefont {G.}~\bibnamefont {Leuchs}}, \ and\ \bibinfo {author}
  {\bibfnamefont {P.}~\bibnamefont {Banzer}},\ }\href
  {https://arxiv.org/pdf/1804.05641.pdf} {\bibfield  {journal} {\bibinfo
  {journal} {arXiv: 1804.05641}\ } (\bibinfo {year} {2018})}\BibitemShut
  {NoStop}%
\bibitem [{\citenamefont {Jackson}(1999)}]{Jackson1999}%
  \BibitemOpen
  \bibfield  {author} {\bibinfo {author} {\bibfnamefont {J.~D.}\ \bibnamefont
  {Jackson}},\ }\href {\doibase 10.4006/1.3025509} {\emph {\bibinfo {title}
  {{Classical Electrodynamics}}}},\ \bibinfo {edition} {3rd}\ ed.\ (\bibinfo
  {publisher} {Wiley},\ \bibinfo {address} {New York},\ \bibinfo {year}
  {1999})\BibitemShut {NoStop}%
\bibitem [{\citenamefont {Bliokh}\ \emph
  {et~al.}(2014{\natexlab{a}})\citenamefont {Bliokh}, \citenamefont {Kivshar},\
  and\ \citenamefont {Nori}}]{Bliokh2014a}%
  \BibitemOpen
  \bibfield  {author} {\bibinfo {author} {\bibfnamefont {K.~Y.}\ \bibnamefont
  {Bliokh}}, \bibinfo {author} {\bibfnamefont {Y.~S.}\ \bibnamefont {Kivshar}},
  \ and\ \bibinfo {author} {\bibfnamefont {F.}~\bibnamefont {Nori}},\ }\href
  {\doibase 10.1103/PhysRevLett.113.033601} {\bibfield  {journal} {\bibinfo
  {journal} {Phys. Rev. Lett.}\ }\textbf {\bibinfo {volume} {113}},\ \bibinfo
  {pages} {033601} (\bibinfo {year} {2014}{\natexlab{a}})}\BibitemShut
  {NoStop}%
\bibitem [{\citenamefont {O'Connor}\ \emph {et~al.}(2014)\citenamefont
  {O'Connor}, \citenamefont {Ginzburg}, \citenamefont
  {Rodr{\'{i}}guez-Fortu{\~{n}}o}, \citenamefont {Wurtz},\ and\ \citenamefont
  {Zayats}}]{OConnor2014}%
  \BibitemOpen
  \bibfield  {author} {\bibinfo {author} {\bibfnamefont {D.}~\bibnamefont
  {O'Connor}}, \bibinfo {author} {\bibfnamefont {P.}~\bibnamefont {Ginzburg}},
  \bibinfo {author} {\bibfnamefont {F.~J.}\ \bibnamefont
  {Rodr{\'{i}}guez-Fortu{\~{n}}o}}, \bibinfo {author} {\bibfnamefont {G.~A.}\
  \bibnamefont {Wurtz}}, \ and\ \bibinfo {author} {\bibfnamefont {A.~V.}\
  \bibnamefont {Zayats}},\ }\href {\doibase 10.1038/ncomms6327} {\bibfield
  {journal} {\bibinfo  {journal} {Nat. Commun.}\ }\textbf {\bibinfo {volume}
  {5}},\ \bibinfo {pages} {5327} (\bibinfo {year} {2014})}\BibitemShut
  {NoStop}%
\bibitem [{\citenamefont {Lukosz}(1979)}]{Lukosz1979}%
  \BibitemOpen
  \bibfield  {author} {\bibinfo {author} {\bibfnamefont {W.}~\bibnamefont
  {Lukosz}},\ }\href {\doibase 10.1364/JOSA.69.001495} {\bibfield  {journal}
  {\bibinfo  {journal} {J. Opt. Soc. Am.}\ }\textbf {\bibinfo {volume} {69}},\
  \bibinfo {pages} {1495} (\bibinfo {year} {1979})}\BibitemShut {NoStop}%
\bibitem [{\citenamefont {Novotny}\ and\ \citenamefont
  {Hecht}(2006)}]{Novotny2006}%
  \BibitemOpen
  \bibfield  {author} {\bibinfo {author} {\bibfnamefont {L.}~\bibnamefont
  {Novotny}}\ and\ \bibinfo {author} {\bibfnamefont {B.}~\bibnamefont
  {Hecht}},\ }\href@noop {} {\emph {\bibinfo {title} {{Principles of
  Nano-Optics}}}},\ \bibinfo {edition} {2nd}\ ed.\ (\bibinfo  {publisher}
  {Cambridge University Press},\ \bibinfo {address} {Cambridge},\ \bibinfo
  {year} {2006})\BibitemShut {NoStop}%
\bibitem [{\citenamefont {Youngworth}\ and\ \citenamefont
  {Brown}(2000)}]{Youngworth2000}%
  \BibitemOpen
  \bibfield  {author} {\bibinfo {author} {\bibfnamefont {K.}~\bibnamefont
  {Youngworth}}\ and\ \bibinfo {author} {\bibfnamefont {T.}~\bibnamefont
  {Brown}},\ }\href {\doibase 10.1364/OE.7.000077} {\bibfield  {journal}
  {\bibinfo  {journal} {Opt. Express}\ }\textbf {\bibinfo {volume} {7}},\
  \bibinfo {pages} {77} (\bibinfo {year} {2000})}\BibitemShut {NoStop}%
\bibitem [{\citenamefont {Alaee}\ \emph {et~al.}(2015)\citenamefont {Alaee},
  \citenamefont {Albooyeh}, \citenamefont {Rahimzadegan}, \citenamefont
  {Mirmoosa}, \citenamefont {Kivshar},\ and\ \citenamefont
  {Rockstuhl}}]{Alaee2015}%
  \BibitemOpen
  \bibfield  {author} {\bibinfo {author} {\bibfnamefont {R.}~\bibnamefont
  {Alaee}}, \bibinfo {author} {\bibfnamefont {M.}~\bibnamefont {Albooyeh}},
  \bibinfo {author} {\bibfnamefont {A.}~\bibnamefont {Rahimzadegan}}, \bibinfo
  {author} {\bibfnamefont {M.~S.}\ \bibnamefont {Mirmoosa}}, \bibinfo {author}
  {\bibfnamefont {Y.~S.}\ \bibnamefont {Kivshar}}, \ and\ \bibinfo {author}
  {\bibfnamefont {C.}~\bibnamefont {Rockstuhl}},\ }\href {\doibase
  10.1103/PhysRevB.92.245130} {\bibfield  {journal} {\bibinfo  {journal} {Phys.
  Rev. B}\ }\textbf {\bibinfo {volume} {92}},\ \bibinfo {pages} {245130}
  (\bibinfo {year} {2015})},\ \Eprint {http://arxiv.org/abs/1508.06965}
  {arXiv:1508.06965} \BibitemShut {NoStop}%
\bibitem [{\citenamefont {Banzer}\ \emph {et~al.}(2010)\citenamefont {Banzer},
  \citenamefont {Peschel}, \citenamefont {Quabis},\ and\ \citenamefont
  {Leuchs}}]{Banzer2010}%
  \BibitemOpen
  \bibfield  {author} {\bibinfo {author} {\bibfnamefont {P.}~\bibnamefont
  {Banzer}}, \bibinfo {author} {\bibfnamefont {U.}~\bibnamefont {Peschel}},
  \bibinfo {author} {\bibfnamefont {S.}~\bibnamefont {Quabis}}, \ and\ \bibinfo
  {author} {\bibfnamefont {G.}~\bibnamefont {Leuchs}},\ }\href {\doibase
  10.1364/OE.18.010905} {\bibfield  {journal} {\bibinfo  {journal} {Opt.
  Express}\ }\textbf {\bibinfo {volume} {18}},\ \bibinfo {pages} {10905}
  (\bibinfo {year} {2010})}\BibitemShut {NoStop}%
\bibitem [{\citenamefont {Marrucci}\ \emph {et~al.}(2006)\citenamefont
  {Marrucci}, \citenamefont {Manzo},\ and\ \citenamefont
  {Paparo}}]{Marrucci2006}%
  \BibitemOpen
  \bibfield  {author} {\bibinfo {author} {\bibfnamefont {L.}~\bibnamefont
  {Marrucci}}, \bibinfo {author} {\bibfnamefont {C.}~\bibnamefont {Manzo}}, \
  and\ \bibinfo {author} {\bibfnamefont {D.}~\bibnamefont {Paparo}},\ }\href
  {\doibase 10.1103/PhysRevLett.96.163905} {\bibfield  {journal} {\bibinfo
  {journal} {Phys. Rev. Lett.}\ }\textbf {\bibinfo {volume} {96}},\ \bibinfo
  {pages} {163905} (\bibinfo {year} {2006})}\BibitemShut {NoStop}%
\bibitem [{\citenamefont {Mick}\ \emph {et~al.}(2014)\citenamefont {Mick},
  \citenamefont {Banzer}, \citenamefont {Christiansen},\ and\ \citenamefont
  {Leuchs}}]{Mick2014}%
  \BibitemOpen
  \bibfield  {author} {\bibinfo {author} {\bibfnamefont {U.}~\bibnamefont
  {Mick}}, \bibinfo {author} {\bibfnamefont {P.}~\bibnamefont {Banzer}},
  \bibinfo {author} {\bibfnamefont {S.}~\bibnamefont {Christiansen}}, \ and\
  \bibinfo {author} {\bibfnamefont {G.}~\bibnamefont {Leuchs}},\ }\href
  {\doibase 10.1364/CLEO_SI.2014.STu1H.1} {\bibfield  {journal} {\bibinfo
  {journal} {Cleo: 2014}\ ,\ \bibinfo {pages} {STu1H.1}} (\bibinfo {year}
  {2014})}\BibitemShut {NoStop}%
\bibitem [{\citenamefont {Schaefer}\ \emph {et~al.}(2007)\citenamefont
  {Schaefer}, \citenamefont {Collett}, \citenamefont {Smyth}, \citenamefont
  {Barrett},\ and\ \citenamefont {Fraher}}]{Schaefer2007}%
  \BibitemOpen
  \bibfield  {author} {\bibinfo {author} {\bibfnamefont {B.}~\bibnamefont
  {Schaefer}}, \bibinfo {author} {\bibfnamefont {E.}~\bibnamefont {Collett}},
  \bibinfo {author} {\bibfnamefont {R.}~\bibnamefont {Smyth}}, \bibinfo
  {author} {\bibfnamefont {D.}~\bibnamefont {Barrett}}, \ and\ \bibinfo
  {author} {\bibfnamefont {B.}~\bibnamefont {Fraher}},\ }\href {\doibase
  10.1119/1.2386162} {\bibfield  {journal} {\bibinfo  {journal} {Am. J. Phys.}\
  }\textbf {\bibinfo {volume} {75}},\ \bibinfo {pages} {163} (\bibinfo {year}
  {2007})}\BibitemShut {NoStop}%
\bibitem [{\citenamefont {Bekshaev}\ \emph {et~al.}(2011)\citenamefont
  {Bekshaev}, \citenamefont {Bliokh},\ and\ \citenamefont
  {Soskin}}]{Bekshaev2011a}%
  \BibitemOpen
  \bibfield  {author} {\bibinfo {author} {\bibfnamefont {A.}~\bibnamefont
  {Bekshaev}}, \bibinfo {author} {\bibfnamefont {K.~Y.}\ \bibnamefont
  {Bliokh}}, \ and\ \bibinfo {author} {\bibfnamefont {M.}~\bibnamefont
  {Soskin}},\ }\href {\doibase 10.1088/2040-8978/13/5/053001} {\bibfield
  {journal} {\bibinfo  {journal} {J. Opt.}\ }\textbf {\bibinfo {volume} {13}},\
  \bibinfo {pages} {053001} (\bibinfo {year} {2011})}\BibitemShut {NoStop}%
\bibitem [{\citenamefont {Dogariu}\ and\ \citenamefont
  {Schwartz}(2006)}]{Dogariu2006}%
  \BibitemOpen
  \bibfield  {author} {\bibinfo {author} {\bibfnamefont {A.}~\bibnamefont
  {Dogariu}}\ and\ \bibinfo {author} {\bibfnamefont {C.}~\bibnamefont
  {Schwartz}},\ }\href {http://www.ncbi.nlm.nih.gov/pubmed/19529220} {\bibfield
   {journal} {\bibinfo  {journal} {Opt. Exp.}\ }\textbf {\bibinfo {volume}
  {14}},\ \bibinfo {pages} {8425} (\bibinfo {year} {2006})}\BibitemShut
  {NoStop}%
\bibitem [{\citenamefont {Bliokh}\ \emph
  {et~al.}(2014{\natexlab{b}})\citenamefont {Bliokh}, \citenamefont {Dressel},\
  and\ \citenamefont {Nori}}]{Bliokh2014b}%
  \BibitemOpen
  \bibfield  {author} {\bibinfo {author} {\bibfnamefont {K.~Y.}\ \bibnamefont
  {Bliokh}}, \bibinfo {author} {\bibfnamefont {J.}~\bibnamefont {Dressel}}, \
  and\ \bibinfo {author} {\bibfnamefont {F.}~\bibnamefont {Nori}},\ }\href
  {\doibase 10.1088/1367-2630/16/9/093037} {\bibfield  {journal} {\bibinfo
  {journal} {New J. Phys.}\ }\textbf {\bibinfo {volume} {16}},\ \bibinfo
  {pages} {093037} (\bibinfo {year} {2014}{\natexlab{b}})},\ \Eprint
  {http://arxiv.org/abs/1404.5486} {arXiv:1404.5486} \BibitemShut {NoStop}%
\bibitem [{\citenamefont {Bliokh}\ \emph {et~al.}(2015)\citenamefont {Bliokh},
  \citenamefont {Rodr{\'{i}}guez-Fortu{\~{n}}o}, \citenamefont {Nori},\ and\
  \citenamefont {Zayats}}]{Bliokh2015}%
  \BibitemOpen
  \bibfield  {author} {\bibinfo {author} {\bibfnamefont {K.~Y.}\ \bibnamefont
  {Bliokh}}, \bibinfo {author} {\bibfnamefont {F.~J.}\ \bibnamefont
  {Rodr{\'{i}}guez-Fortu{\~{n}}o}}, \bibinfo {author} {\bibfnamefont
  {F.}~\bibnamefont {Nori}}, \ and\ \bibinfo {author} {\bibfnamefont {A.~V.}\
  \bibnamefont {Zayats}},\ }\href {\doibase 10.1038/nphoton.2015.201}
  {\bibfield  {journal} {\bibinfo  {journal} {Nat. Photon.}\ }\textbf {\bibinfo
  {volume} {9}},\ \bibinfo {pages} {796} (\bibinfo {year} {2015})}\BibitemShut
  {NoStop}%
\bibitem [{\citenamefont {Takeda}\ \emph {et~al.}(1982)\citenamefont {Takeda},
  \citenamefont {Ina},\ and\ \citenamefont {Kobayashi}}]{Takeda1982}%
  \BibitemOpen
  \bibfield  {author} {\bibinfo {author} {\bibfnamefont {M.}~\bibnamefont
  {Takeda}}, \bibinfo {author} {\bibfnamefont {H.}~\bibnamefont {Ina}}, \ and\
  \bibinfo {author} {\bibfnamefont {S.}~\bibnamefont {Kobayashi}},\ }\href
  {\doibase 10.1364/JOSA.72.000156} {\bibfield  {journal} {\bibinfo  {journal}
  {J. Opt. Soc. Am.}\ }\textbf {\bibinfo {volume} {72}},\ \bibinfo {pages}
  {156} (\bibinfo {year} {1982})}\BibitemShut {NoStop}%
\end{thebibliography}%
\end{document}